\documentclass[twocolumn]{aastex62}

\begin{document}

\title{\bf \large The Predicament of Absorption-Dominated Reionization:\\ Increased Demands on Ionizing Sources}
\shorttitle{Absorption-dominated Reionization}
\shortauthors{F. B. Davies et al.}

\correspondingauthor{Frederick B.~Davies}
\email{davies@mpia.de}

\author[0000-0003-0821-3644]{Frederick B.~Davies}
\affiliation{Max-Planck-Institut f\"{u}r Astronomie, K\"{o}nigstuhl 17, D-69117 Heidelberg, Germany}

\author[0000-0001-8582-7012]{Sarah E.~I.~Bosman}
\affiliation{Max-Planck-Institut f\"{u}r Astronomie, K\"{o}nigstuhl 17, D-69117 Heidelberg, Germany}

\author[0000-0002-0658-1243]{Steven R.~Furlanetto}
\affiliation{Department of Physics \& Astronomy, University of California, Los Angeles, CA 90095, USA}

\author[0000-0003-2344-263X]{George D.~Becker}
\affiliation{Department of Physics \& Astronomy, University of California, Riverside, CA 92521, USA}

\author{Anson D'Aloisio}
\affiliation{Department of Physics \& Astronomy, University of California, Riverside, CA 92521, USA}

\defcitealias{Becker21}{B21}

\begin{abstract}
The reionization epoch concludes when ionizing photons reach every corner of the Universe. Reionization has generally been assumed to be limited primarily by the rate at which galaxies produce ionizing photons, but the recent measurement of a surprisingly short ionizing photon mean free path of $0.75^{+0.65}_{-0.45}$ proper Mpc at $z = 6$ by \citet{Becker21} suggests that absorption by residual neutral hydrogen in the otherwise ionized intergalactic medium may play a much larger role than previously expected. Here we show that consistency between this short mean free path and the coeval dark pixel fraction in the Ly$\alpha$ forest requires a cumulative output of $6.1^{+11}_{-2.4}$ ionizing photons per baryon by reionization's end. This represents a dramatic increase in the ionizing photon budget over previous estimates, greatly exacerbating the tension with measurements of the ionizing output from galaxies at later times. Translating this constraint into the instantaneous ionizing production from galaxies in our model, we find$\log_{10}f_{\rm esc}\xi_{\rm ion}/\text{(erg/Hz)}^{-1} =25.02_{-0.21}^{+0.45}$ at $z\sim6$. Even with optimistic assumptions about the ionizing production efficiency of early stellar populations, and assuming the galaxy luminosity function extends to extremely faint sources ($M_{\text{UV}}\leq-11$), complete reionization requires the escape fraction of ionizing photons to exceed $20\%$ across the galaxy population. This is far larger than observed in any galaxy population at lower redshifts, requiring rapid evolution in galaxy properties after the first billion years of cosmic time. This tension cannot be completely relieved within existing observational constraints on the hydrogen neutral fraction and mean free path.
\end{abstract}


\section{Introduction}

The epoch of reionization is one of the landmark events in the early history of galaxy formation, during which hydrogen in the intergalactic medium (IGM) -- which had been neutral since cosmological recombination at $z \sim 1100$ -- was ionized by early galaxies. The measured transmission of Ly$\alpha$ photons through the IGM suggests that reionization was complete by $z\sim5$--$6$ \citep{Bosman18,Eilers18,Yang20b}, but much is still unknown about the process.

The reionization history is fundamentally connected to the evolution of the sources and sinks of ionizing photons across cosmic time. The sources are thought to be star-forming galaxies (e.g.~\citealt{Robertson15}), whose populations have been catalogued via their (non-ionizing) UV luminosity functions (LFs) up to $z\sim10$ (e.g.~\citealt{Bouwens21}). 
The sinks are thought to be dense clumps of IGM gas that can shield themselves from ionizing radiation and remain neutral \citep{MHR00} even after their volume has been reionized. These clumps manifest as Lyman-limit systems, and their cumulative opacity defines the ``mean free path'' $\lambda$ of ionizing photons in the later Universe (e.g.~\citealt{SC10}).
The mean free path can be measured directly via the average attenuation observed in stacked 
quasar spectra \citep{Prochaska09,Worseck14}. 

In most reionization models, the absorption of ionizing photons is assumed to be modest, requiring only $\sim1$--$3$ ionizing photons per baryon (e.g.~\citealt{Gnedin08,Finlator12}). In that case, the observed LF 
evolution is consistent with reionization completing at $z \sim 6$, provided the escape fraction of ionizing photons, $f_{\rm esc}$ from these galaxies is somewhat larger ($\ga 10\%$) \citep{Robertson15} than that measured directly in later galaxies (e.g.~\citealt{Pahl21}).

\begin{figure*}[ht]
\begin{center}
\resizebox{18cm}{!}{\includegraphics[trim={1em 1em 1em 0em},clip]{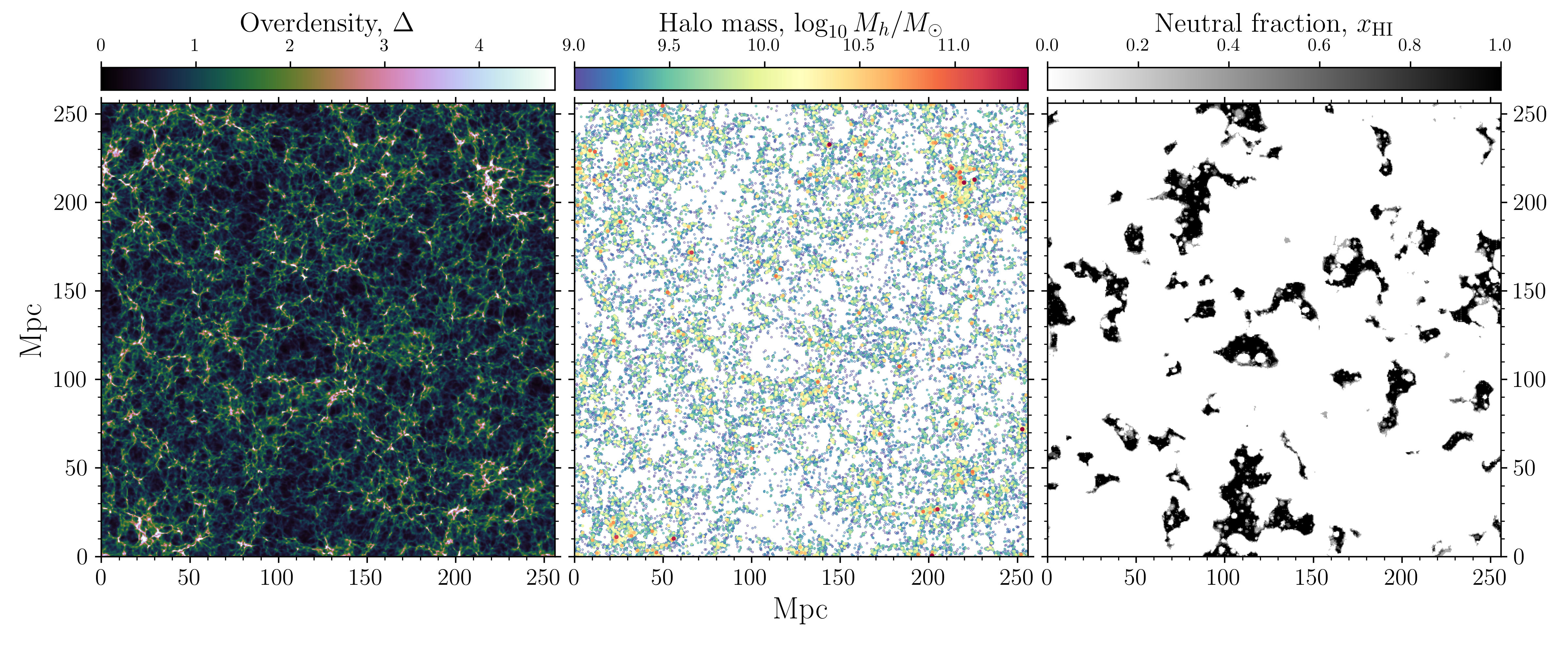}}\\
\end{center}
\caption{Slice (1.5\,Mpc-thick) through our fiducial 
simulation with $L_{\rm box}=256$\,Mpc, $\lambda=5.25$\,Mpc, and $M_{\rm min}=10^9\,M_\odot$, matched to the dark pixel fraction $84\%$ confidence upper limit on the volume-fraction of neutral hydrogen at $z=5.9$ from \citet{McGreer15}. Left: matter overdensity. Middle: dark matter halo distribution, color- and size-coded by mass. Right: hydrogen neutral fraction.}
\label{fig:slices}
\end{figure*}

Recently, the mean free path at $z\sim6$ has been measured by \citet[][henceforth \citetalias{Becker21}]{Becker21} to be $\lambda = 0.75^{+0.65}_{-0.45}$ proper Mpc ($5.25^{+4.55}_{-3.15}$\, comoving Mpc), far shorter than the extrapolation from previous measurements at $z\lesssim5$ using a similar method \citep{Worseck14}. In this Letter, we investigate the implications of this short mean free path for the reionization epoch using the semi-numerical method of \citealt{DF21}. We find that the short mean free path implies a dramatic increase in the number of emitted ionizing photons required to reionize the Universe. This in turn requires a substantial increase in the ionizing efficiency of galaxies at $z>6$ such that their intrinsic properties differ significantly from lower-redshift galaxies.

We adopt a \citet{Planck18} cosmology with $(h,\Omega_m,\Omega_\Lambda,\Omega_b,\sigma_8,n_s)$ $= (0.6736, 0.3153, 0.6847,$ $0.0493, 0.8111, 0.9649)$, and use comoving distances unless specified otherwise.

\section{Simulation Method}

To simulate the effect of a short mean free path on reionization, we use the semi-numerical method of \citet{DF21}, consisting of a modified version of the \texttt{21cmFAST} code \citep{Mesinger11}. The simulations start with cosmological initial conditions (ICs) in a volume of size $L_{\rm box}$ with $N_{\rm IC}$ cells on a side. We then evolve the density field via the Zel'dovich approximation \citep{Zel'dovich70} at a coarsened resolution with $N_{\rm ev}$ cells on a side. We employ the halo-filtering methodology of \citet{MF07} to identify dark matter halos and evolve their positions using the Zel'dovich displacement field. Our fiducial simulations adopt $L_{\rm box}=256$\,Mpc, $N_{\rm IC}=4096$, and $N_{\rm ev}=512$, 
so that $L_{\rm cell}=0.5$\,Mpc. We explore a range of $L_{\rm box}$ from $128$ to $512$ Mpc at fixed $N_{\rm IC}=4096$ and evolved $L_{\rm cell}=0.5$\,Mpc\footnote{We find little difference for $L_{\rm cell}=0.25$ or $1.0$\,Mpc.}. To limit numerical noise in the halo distribution we adopt a minimum halo mass $M_{\rm min}$ corresponding to $\sim100$ mean density IC cells, so each $L_{\rm box}$ has a corresponding $M_{\rm min}$. 

\begin{figure*}[ht]
\begin{center}
\resizebox{18cm}{!}{\includegraphics[trim={1em 1em 1em 1em},clip]{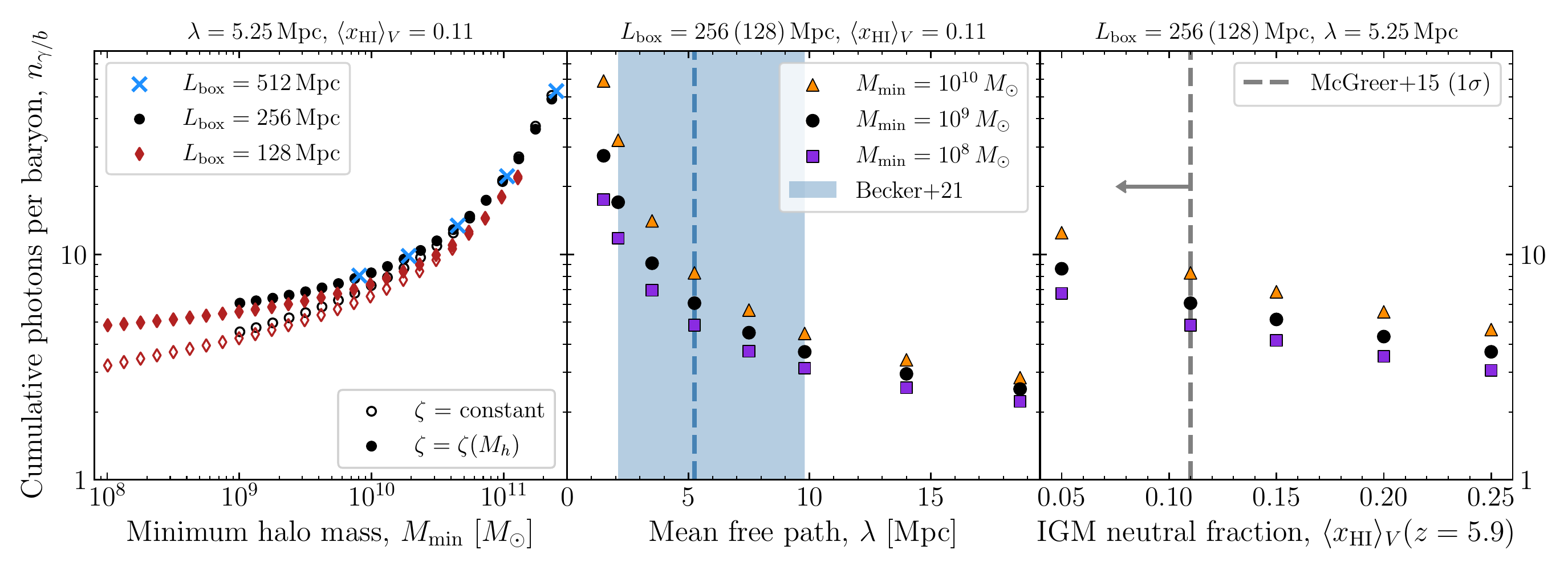}}\\
\end{center}
\caption{Number of ionizing photons per baryon required to reionize the Universe in our simulations (points). Left: dependence on minimum halo mass, simulation box size, and $\zeta$ model. 
Middle: dependence on mean free path. Due to resolution limitations, simulations with $M_{\rm min}=10^8\,M_\odot$ have $L_{\rm box}=128$\,Mpc. The \citetalias{Becker21} mean free path constraint is shown by the blue shaded region. Right: dependence on the volume-averaged neutral fraction at $z=5.9$. The \citet{McGreer15} 84\% confidence upper limit is shown by the vertical dashed line.}
\label{fig:npb}\label{fig:budget}
\end{figure*}

We assume that the mass of material ionized by a halo, in the absence of absorption, is related to its mass via 
\begin{equation}\label{eqn:zeta}
    M_{\rm ion} = \zeta M_h,
\end{equation}
This $\zeta$ is the product of several parameters of stars in early galaxies, 
\begin{equation} \label{eqn:zeta2}
\zeta=f_{\rm esc} f_* N^*_{\gamma/b},
\end{equation}
where $f_*$ is the fraction of the halo's baryons which formed stars and $N^*_{\gamma/b}$ is the (integrated) number of ionizing photons emitted per stellar baryon \citep{Furlanetto04}.

We employ two different prescriptions for $\zeta$.The first assumes that $\zeta=\zeta_0$ is constant as a function of halo mass, i.e.~$M_{\rm ion}\propto M_h$. This model is motivated by simplicity and by its use in many past works. Our fiducial prescription for $\zeta$ assumes a double power-law functional form,
\begin{equation}\label{eqn:dpl}
    \zeta(M_h) = \zeta_0 \frac{(M_0/M_{\rm peak})^{-\gamma_{\rm lo}}+(M_0/M_{\rm peak})^{-\gamma_{\rm hi}}}{(M_h/M_{\rm peak})^{-\gamma_{\rm lo}}+(M_h/M_{\rm peak})^{-\gamma_{\rm hi}}},
\end{equation}
where $\zeta_0$ is a normalization factor applied at a halo mass $M_0$, $\gamma_{\rm lo}=0.49$ and $\gamma_{\rm hi}=-0.61$ are effective power-law indices at low and high mass, respectively, and $M_{\rm peak}=2.8\times10^{11}\,M_\odot$ is a transition mass between the two.
This form (and the corresponding parameters) were originally chosen by \citet{Mirocha17} to characterize the relationship between star formation and halo accretion required to reproduce observed UV LFs at $z=6$--$8$ (see also \citealt{Mirocha20b}). Note that we apply this functional form to the ratio between the \emph{total} number of stars formed and the halo \emph{mass} but in practice the shapes of the two relations should be very similar \citep{Furlanetto17}. The shape can be interpreted as accounting for a suppression of star formation at low halo masses due to supernova feedback, and at high halo masses due slow gas cooling or AGN feedback \citep{Furlanetto17}. 

We use the ``MFP-$\epsilon(r)$'' approach from \citet{DF21} to compute the ionization topology in the presence of absorption. 
The standard framework for semi-numerical reionization simulations relies on a photon-counting argument, whereby a region is ionized if the number of ionizing photons produced inside it exceeds the number of hydrogen atoms. This argument is typically applied to a single cell at the center of the region (e.g.~\citealt{Zahn11}). \citet{DF21} showed that this central cell calculation is equivalent to integrating the ionizing flux, enabling distance-dependent absorption to be efficiently included. In the MFP-$\epsilon(r)$ approach, the ionization criterion on scale $R$ is given by
\begin{equation}
    \int_0^\infty \langle \zeta f_{\rm coll} \Delta \rangle_r W(r; R,\lambda)\,4\pi r^2dr > \langle \Delta(<R) \rangle,
\end{equation}
where $\Delta$ is the overdensity relative to the cosmic mean, $f_{\rm coll}$ is the fraction of matter in halos above the minimum mass for galaxies to form, $\langle \zeta f_{\rm coll} \Delta \rangle_r$ is averaged within the spherical shell at radius $r$, and $W(r; R,\lambda)$ is a spherical top-hat filter of size $R$ in real space with an exponential attenuation factor $e^{-r/\lambda}$ to account for absorption in the ionized gas. In Figure~\ref{fig:slices}, we show an example of the evolved density field, halo distribution, and neutral fraction field of our fiducial simulation.

Finally, we note that semi-numerical simulations that use a real-space filter to compute the ionization topology are known to suffer from photon non-conservation (e.g. \citealt{Zahn07}) at the $\lesssim20\%$ level, peaking when ionized bubbles first overlap and weakening towards the later stages of reionization \citep{Hutter18}. We expect a similar effect to be present in our adopted method, but we note that the uncertainty in the observed mean free path is far larger than $20\%$ and so defer a quantitative comparison to radiative transfer simulations to future work.

\section{The Ionizing Photon Budget}

We aim to constrain the number of ionizing photons that must have been emitted during reionization given the short $z\sim6$ mean free path measured by \citetalias{Becker21}. We make the simplifying assumption that the mean free path prior to $z=6$ is constant; in reality, it is likely shorter at earlier times, requiring even more ionizing photons to complete the process. We then require that the IGM neutral fraction at $z\sim6$ be consistent with the Ly$\alpha$ dark pixel fraction measurement of \citet{McGreer15}, who found a volume-averaged neutral fraction $\langle x_{\rm HI} \rangle_V(z=5.9) < 0.11$ at 84\% confidence. We thus tune $\zeta_0$ for each simulation to match $\langle x_{\rm HI} \rangle_V(z=5.9) = 0.11$, and also explore variations of this value. The resulting mass-averaged IGM neutral fraction in the simulations is $\langle x_{\rm HI}\rangle_M\sim6\%$. In this work we conservatively consider reionization to be ``completed'' at this point -- however, vestiges of incomplete reionization may be detectable at later times in the Ly$\alpha$ forest (\citealt{Kulkarni19,Keating19, ND20}, Bosman et al.~in prep.) or in future 21 cm measurements \citep{Raste21}. 

We quantify the ionizing photon budget by the cumulative number of ionizing photons per baryon,
\begin{equation}
    n_{\gamma/b} = \frac{n_{\rm ion}}{\bar{n}_b} = \langle \zeta f_{\rm coll} \rangle,
\end{equation}
where $n_{\rm ion}$ is the number density of all ionizing photons ever emitted by galaxies and $\bar{n}_b$ is the mean baryon density. The product $\langle \zeta f_{\rm coll}\rangle$ represents the average over any halo mass dependence of $\zeta$. Our fiducial simulation with $L_{\rm box}=256$\,Mpc, $M_{\rm min}=10^9\,M_\odot$, $\langle x_{\rm HI}\rangle_V(z=5.9)=0.11$, and the double power-law $\zeta$ model gives $n_{\gamma/b}=6.1^{+11}_{-2.4}$, with uncertainty from the \citetalias{Becker21} mean free path measurement alone. The central value is $\sim 2$--$6$ times larger than typical estimates from earlier models (e.g.~\citealt{Gnedin08,Finlator12}).

In Figure~\ref{fig:budget} we show how $n_{\gamma/b}$ varies across the models in our simulation suite. The left panel shows $n_{\gamma/b}$ as a function of $M_{\rm min}$ for the constant $\zeta$ models (open symbols) and fiducial double power-law $\zeta$ models (solid symbols) for different box sizes; in the smaller (larger) simulations we can resolve less (more) massive halos. Emphasizing lower mass halos, i.e.~by decreasing $M_{\rm min}$ or treating $\zeta$ as constant, generally requires fewer photons per baryon to complete reionization, because low-mass galaxies are less biased and therefore spatially closer to the voids that are ionized during the later phases of reionization. Their proximity decreases the attenuation from the short mean free path. Meanwhile, the large-scale voids corresponding to the final $\sim10\%$ ionization of the Universe represent rare, large, underdense structures, so the required photon budget could depend on features on the largest scales. Convergence is achieved for $L_{\rm box}\gtrsim200$\,Mpc, motivating our focus on a fiducial $L_{\rm box}=256$\,Mpc simulation suite. 

The middle panel of Figure~\ref{fig:budget} shows the dependence of $n_{\gamma/b}$ on the mean free path. This trend is the dominant one in our simulations, reflecting the importance of the mean free path for the end stages of reionization. The right panel shows the dependence of $n_{\gamma/b}$ on the assumed IGM neutral fraction at $z=5.9$. If the neutral fraction at $z=5.9$ lies significantly below the \citet{McGreer15} upper limit of $0.11$, the required number of photons would increase substantially.

\begin{figure}[t]
\begin{center}
\resizebox{7.5cm}{!}{\includegraphics[trim={1em 1em 1em 1em},clip]{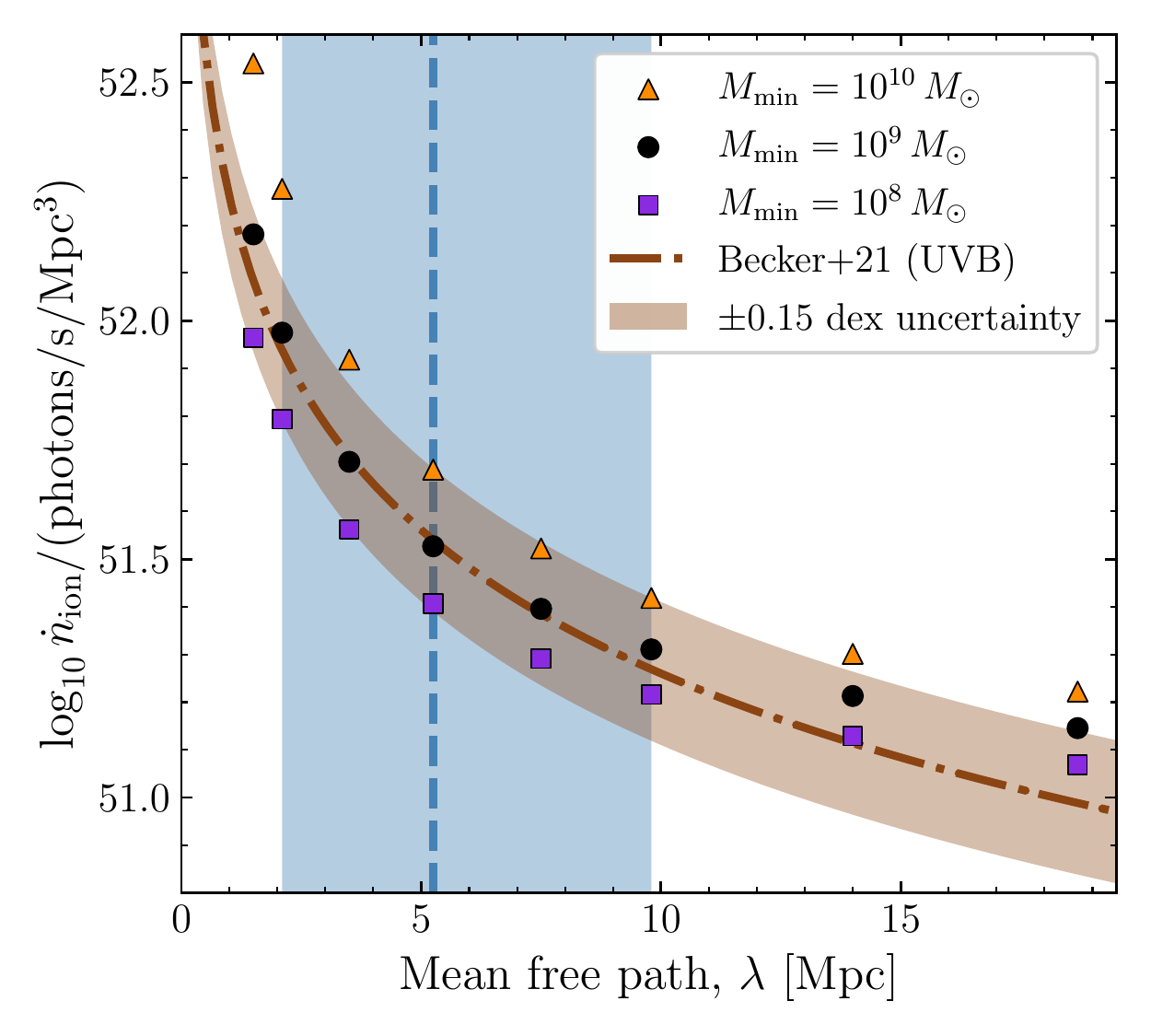}}\\
\end{center}
\caption{Ionizing emissivity at $z=5.9$ in our simulations computed via equation~(\ref{eqn:ndot_sim}) as a function of mean free path. The brown dot-dashed curve shows the emissivity that reproduces the mean UV background estimated by \citetalias{Becker21}, with their $\Gamma_{\rm HI}$ uncertainty shown by the shaded region.}
\label{fig:nion}
\end{figure}

\section{Implications for Reionization-Epoch Galaxies}\label{sec:gal}

\begin{figure*}[ht]
\begin{center}
\resizebox{18cm}{!}{\includegraphics[trim={1em 1em 1em 1em},clip]{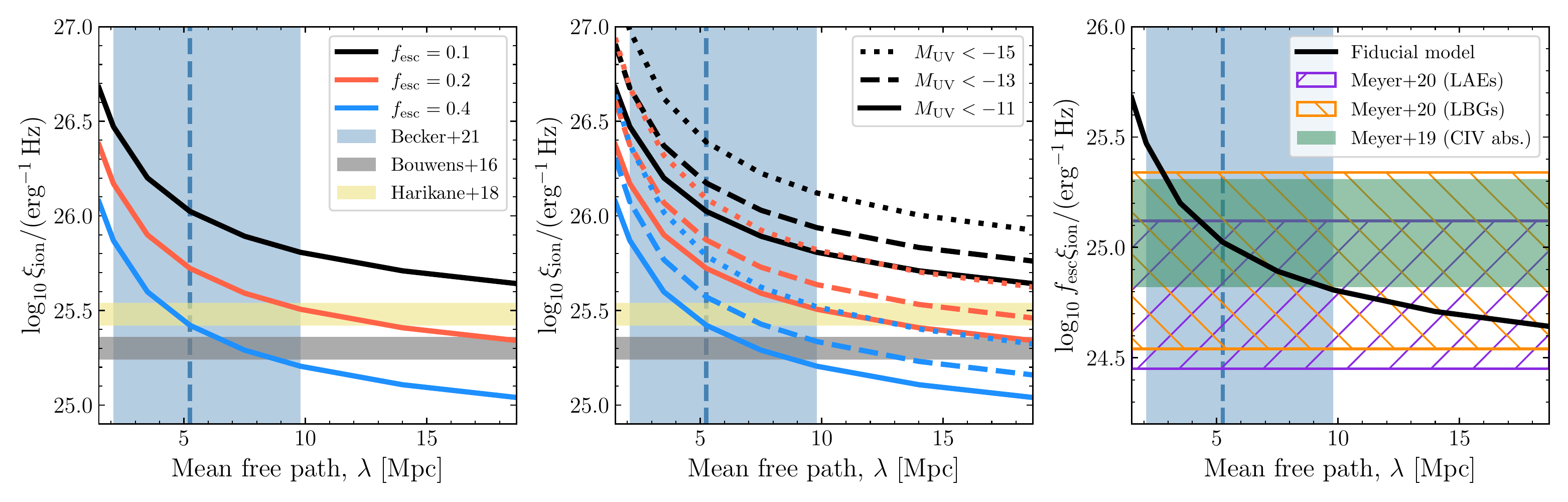}}\\
\end{center}
\caption{Ionizing photon production and escape (at $z=5.9$) as a function of mean free path. Left: curves show $\xi_{\rm ion}$ in the fiducial model for varying $f_{\rm esc}$ compared to measurements at $z\sim4$--$5$ \citep{Bouwens16,Harikane18} shown by horizontal bands. Middle: similar to the left panel, but integrating down to different $M_{\rm UV}$. Right: the black curve shows $f_{\rm esc}\xi_{\rm ion}$ in the fiducial model compared to Ly$\alpha$ forest cross-correlation measurements at $z\sim5$--$6$ \citep{Meyer19,Meyer20} shown by the horizontal bands and hashes.}
\label{fig:fesc_xi}
\end{figure*}

While the cumulative number of photons per baryon $n_{\gamma/b}$ (and its corresponding $\zeta$ prescription) provides a useful cosmological benchmark for reionization studies, it is a time-integrated quantity, which is not trivially related to the galaxy population at any given time. Here we translate our constraint on $n_{\gamma/b}$ to more familiar galaxy properties.

First, we estimate the ionizing emissivity $\dot{n}_{\rm ion}$ from our simulations via differentiating equation~(\ref{eqn:zeta}),
\begin{equation}\label{eqn:ndot_sim}
    \frac{dM_{\rm ion}}{dt} = \frac{d}{dt}(\zeta M_h) = \zeta \frac{dM_h}{dt} + M_h \frac{d\zeta}{dM_h}\frac{dM_h}{dt},
\end{equation}
where $d\zeta/dM_h$ is obtained by differentiating equation~(\ref{eqn:dpl}). We then sum the contribution from every halo $i$ in the simulation and divide by volume to recover $\dot{n}_{\rm ion}$,
\begin{equation}
\dot{n}_{\rm ion} = \frac{1}{L_{\rm box}^3}\sum_i \frac{(\Omega_b/\Omega_m)\dot{M}_{{\rm ion},i}}{\mu m_p},
\end{equation}
where $\Omega_b/\Omega_m$ is the cosmic baryon fraction and $\mu m_p$ is the average baryon mass. We estimate the halo accretion rate $dM_h/dt$ via the abundance matching-like procedure described in \citet{Furlanetto17}, wherein dark matter halos grow roughly exponentially in agreement with cosmological simulations \citep{Dekel13}.

In Figure~\ref{fig:nion} we show the resulting $\dot{n}_{\rm ion}$ at $z=5.9$ as a function of mean free path. Our fiducial model ($M_{\rm min}=10^9\,M_\odot$, $\langle x_{\rm HI} \rangle_V(z=5.9)=0.11$) has $\dot{n}_{\rm ion} =  3.4^{+6.1}_{-1.3}\times10^{51}$\,photons/s/Mpc$^{3}$, where the uncertainties reflect the $1\sigma$ range of $\lambda$. The ionizing emissivity can also be estimated independently by requiring that the photoionization rate in the IGM, $\Gamma_{\rm HI}\propto \dot{n}_{\rm ion} \lambda$, reproduces the observed mean transmission in the Ly$\alpha$ forest. We show $\dot{n}_{\rm ion}$ determined this way by \citetalias{Becker21} as the brown dot-dashed curve in Figure~\ref{fig:nion}. The two $\dot{n}_{\rm ion}$ estimates agree remarkably well across the $1\sigma$ range of the \citetalias{Becker21} mean free path measurement, although this agreement may be somewhat accidental, as our halo-galaxy mapping is approximate and the UV background measurement may be somewhat biased by fluctuations present at $z\sim6$ \citep{DF16,Davies17b,Becker18}.

\begin{figure*}[ht]
\begin{center}
\resizebox{17cm}{!}{\includegraphics[trim={1em 1em 1em 1em},clip]{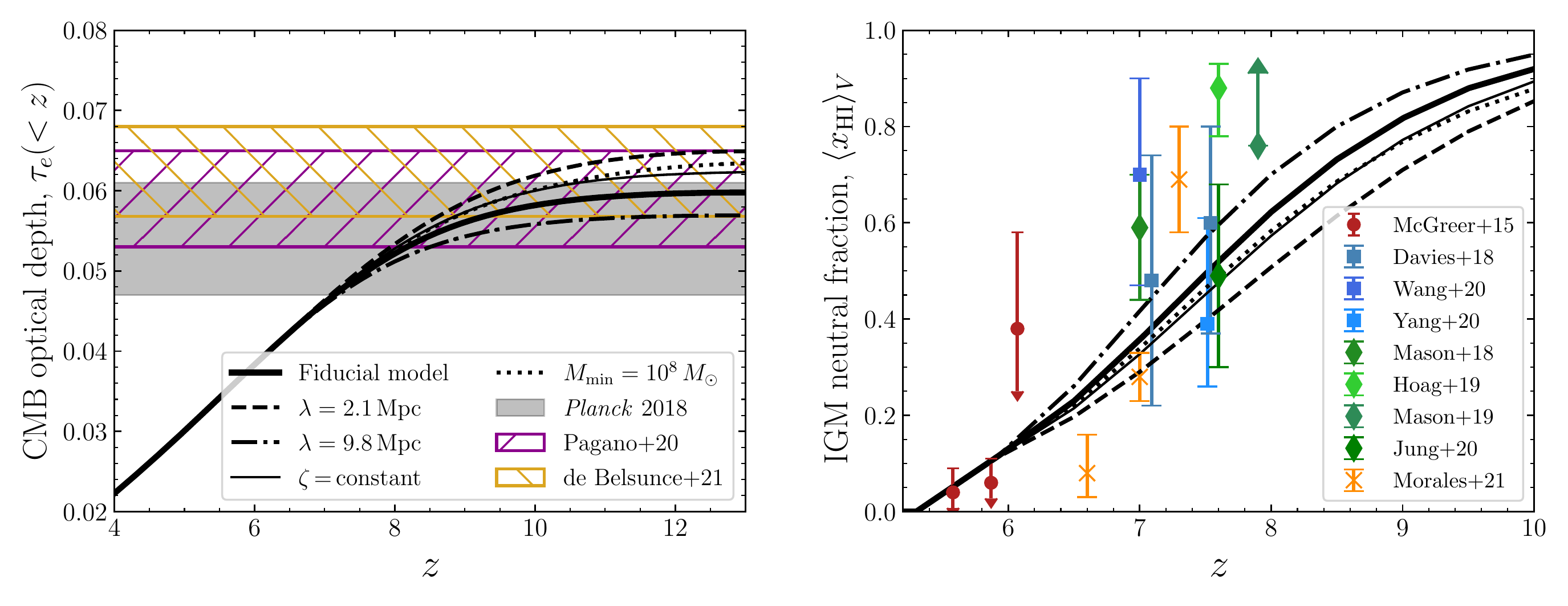}}\\
\end{center}
\caption{Consistency of our model with observational constraints on reionization. Note that we have not simulated the $z<5.9$ evolution, but instead assume the increasing mean free path completes reionization by $z=5.3$. Left: curves show $\tau_{\rm e}$ in the fiducial model (thick solid), $\lambda\pm1\sigma$ (dashed and dot-dashed), constant $\zeta$ (thin solid), and $M_{\rm min}=10^8\,M_\odot$ (dotted). Constraints are shown from \citet{Planck18} (grey shaded), \citet{Pagano20} (purple hash), and \citet{deBelsunce21} (yellow hash). Right: evolution of the volume-averaged neutral fraction. Blue squares show constraints from quasar damping wings \citep{Davies18b,Wang20,Yang20a}, green diamonds show constraints from Ly$\alpha$ and UV emission from galaxies \citep{Mason18,Mason19,Hoag19,Jung20}, and orange crosses show a recent constraint from the evolution of the LAE LF \citep{Morales21}.}
\label{fig:reion}
\end{figure*}

The ionizing emissivity is typically parameterized by
\begin{equation}
\dot{n}_{\rm ion} = f_{\rm esc} \xi_{\rm ion} \rho_{\rm UV},
\end{equation}
where $\xi_{\rm ion}$ is the ``ionizing efficiency'' of the stellar populations, dependent on their metallicity and star formation history, and $\rho_{\rm UV}$ is the integrated UV luminosity density. From $\dot{n}_{\rm ion}$ we can thus estimate the product $f_{\rm esc} \xi_{\rm ion}$ via observational constraints on $\rho_{\rm UV}$ at $z\sim6$. Adopting the best-fit star formation efficiency as a function of halo mass from \citet{Mirocha17}, we estimate that our fiducial $M_{\rm min}=10^9\,M_\odot$ corresponds to $M_{\rm UV}\sim-11$. We then obtain $\rho_{\rm UV}=3.2 \times 10^{26}$ erg/s/Hz/Mpc$^{3}$ by integrating the $z\sim6$ UV LF from \citet{Bouwens21} down to $M_{\rm UV}=-11$. The solid curve in the right panel of Figure~\ref{fig:fesc_xi} shows the resulting $f_{\rm esc} \xi_{\rm ion}$ from our fiducial model. Taking into account the mean free path uncertainty alone, we estimate $\log_{10} f_{\rm esc} \xi_{\rm ion} /\text{(erg/Hz)}^{-1} =25.02_{-0.21}^{+0.45}$. 

Assuming $f_{\rm esc} = 0.1$, this corresponds to an ionizing efficiency of $\log_{10} \xi_{\rm ion}/\text{(erg/Hz)}^{-1} = 26.02_{-0.21}^{+0.45}$, considerably larger than previous measurements of $\log_{10} \xi_{\rm ion}/\text{(erg/Hz)}^{-1} =25.24-25.36$ for $3.8<z<5$ galaxies (\citealt{Bouwens16}), $\log_{10} \xi_{\rm ion}/\text{(erg/Hz)}^{-1}=25.48\pm0.06$ for $z\sim4.9$ Ly$\alpha$-emitting galaxies (LAEs; \citealt{Harikane18}), and even rare, extremely blue galaxies with $\log_{10} \xi_{\rm ion}/\text{(erg/Hz)}^{-1} \sim 25.8$ (\citealt{Bouwens16}; see also \citealt{Stark17}). The left panel of Figure~\ref{fig:fesc_xi} shows that maintaining consistency with either the canonical $\log_{10} \xi_{\rm ion}/\text{(erg/Hz)}^{-1} \sim 25.2$ or the measured value of $25.5$ at $z\sim5$ requires $f_{\rm esc} \gtrsim 30\%$ or $\gtrsim20\%$, respectively. Note that, as mentioned above, these constraints reflect contributions from halos down to $M_{\rm min}=10^9\,M_\odot$, corresponding to $M_{\rm UV}<-11$. The middle panel of Figure~\ref{fig:fesc_xi} shows that 
increasing the minimum luminosity of contributing galaxies, where $M_{\rm UV}=-13$ $(-15)$ corresponds to $M_{\rm min}=10^{9.5}$ $(10^{10})\,M_\odot$, further increases the required $f_{\rm esc} \xi_{\rm ion}$. 

Our values of $f_{\rm esc} \xi_{\rm ion}$ are consistent with \citet{Meyer19,Meyer20} (see also \citealt{Kakiichi18}) who measured the ionizing production of $z\gtrsim5$ galaxies via their correlation with Ly$\alpha$ transmission in background quasar spectra. In the right panel of Figure~\ref{fig:fesc_xi}, we compare $f_{\rm esc} \xi_{\rm ion}$ in our model to values of
$f_{\rm esc} \xi_{\rm ion}$ values measured from three independent populations of galaxies: Lyman-break galaxies (LBGs), LAEs, and the hosts of \ion{C}{4} absorption systems. We note, however, that the measurements from LAEs and LBGs rely non-trivially on much larger assumed mean free paths at $z \sim 6$.

Finally, we demonstrate the consistency of our results with measurements of the stellar mass function. We can estimate $\xi_{\rm ion}$ directly from $\zeta$ using $f_*(M_h)$ from \citet{Mirocha17} and estimating $N^*_{\gamma/b}$ using the integral constraint,
\begin{equation}\label{eqn:nspb}
N^*_{\gamma/b} = \frac{\mathcal{E}_{\text{tot}}|_{z\geq6}}{M_{*,\text{tot}} |_{z=6}} \xi_{\text{ion}},
\end{equation}
where $\mathcal{E}_{\text{tot}}|_{z\geq6}$ is the energy density in UV photons emitted before $z=6$, i.e.
\begin{equation}
    \mathcal{E}_{\text{tot}}|_{z\geq6} = \int_{t(z=6)} \int_{L_{\text{min}}} L \Phi(L,t) dL dt
\end{equation}
and $M_{*,\text{tot}} |_{z=6}$ is the stellar mass density at $z=6$. Using the best-fit UV LFs from \citet{Bouwens21} at $z=5.9, 6.8, 7.9, 8.9$ and \citet{Oesch18} at $z=10.2$, we obtain $\mathcal{E}_{\text{tot}}|_{z\geq6} =3.9 \times 10^{43}$ erg/Hz/Mpc$^3$, integrating down to our nominal halo mass threshold of $10^9\,M_\odot$. Dividing our star formation efficiency model by a factor of $1+\gamma_{\rm lo}$ to approximately recover the stellar mass-halo mass relation \citep{Furlanetto17}, this mass threshold corresponds to a minimum stellar mass of $M_{*,{\rm min}}\sim6.6\times10^5\,M_\odot$, implying a mass-to-light ratio of $\sim0.12 M_\odot/L_\odot$ at $M_{*,{\rm min}}$. This mass-to-light ratio is consistent with measurements in (much brighter) galaxies at $z\geq 6$ (e.g.~\citealt{McLure11}). Equations~(\ref{eqn:zeta2}) and (\ref{eqn:nspb}) then require $M_{*,\text{tot}} |_{z=6} = 8.8 \times 10^6 M_\odot$/Mpc$^3$ to match our constraints on $f_{\rm esc} \xi_{\rm ion}$, which falls squarely between faint-end extrapolations of the stellar mass functions of \citet{Bhatawdekar19} ($M_{*,\text{tot}} |_{z=6} = 1.1 \times 10^7 M_\odot$/Mpc$^3$) and \citet{Song16} ($M_{*,\text{tot}} |_{z=6} = 4.6 \times 10^6 M_\odot$/Mpc$^3$).
Thus our constraints are not in tension with extrapolations of the $z\geq6$ luminosity and stellar mass functions.

We therefore find that reionization-era galaxies must be producing or leaking ionizing photons with efficiencies a factor $2-3$ times higher than even the most optimistic observational estimates to match the $\langle x_{\text{HI}}\rangle_V$ constraint at $z=5.9$. The tension could be eased if the UV LF is much steeper at $z\sim6$, e.g.~with a faint-end slope at the $2\sigma$ limit from \citet{Bouwens21} the required $f_{\rm esc} \xi_{\rm ion}$ would decrease by a factor of two. A strongly declining $f_{\rm esc}$ with halo mass could also decrease the required photon budget; the constant $\zeta$ model, implying $f_{\rm esc}\propto M_h^{-0.49}$, decreases $n_{\gamma/b}$ by $\sim 30\%$. \citet{Cain21} have similarly explored the ionizing budget requirements of reionization with a short mean free path using radiative transfer simulations that include a simulation-calibrated sub-grid model for small-scale gas clumping. Their fiducial simulation assumes that every halo has the same ionizing luminosity, placing even more emphasis on the faintest galaxies. Their ionizing photon budget at $z=5.9$ of $n_{\gamma/b}\sim2.2$ is nevertheless within $\sim10\%$ of our constant $\zeta$ model when run with values of $\langle x_{\rm HI}\rangle_V=0.2$ and $\lambda=10$\,Mpc similar to their model (at the $\sim+2\sigma$ and $\sim+1\sigma$ of current observations, respectively), which gives $n_{\gamma/b}\approx2.3$.

\newpage
\section{Conclusion}

In this work, we have explored the consequences of a short mean free path of ionizing photons (as measured by \citetalias{Becker21}) on the ionizing photon budget of the reionization epoch. Using the new semi-numerical method of \citet{DF21}, we found that their mean free path of $0.75^{+0.65}_{-0.45}$ proper Mpc implies a cumulative ionizing photon budget of $6.1^{+11}_{-2.4}$ photons per baryon to reproduce the $1\sigma$ upper limit on the $z=5.9$ IGM neutral fraction of $\langle x_{\rm HI} \rangle_V<0.11$ from \citet{McGreer15}. The large number of required photons is linked to the distance that the photons must travel from sources to the last remaining neutral voids. 

We translated our constraint on the cumulative number of photons per baryon into the instantaneous emission properties of reionization-epoch galaxies, finding an implied ionizing production rate of $\log_{10} f_{\rm esc} \xi_{\rm ion} /\text{(erg/Hz)}^{-1} =25.02_{-0.21}^{+0.45}$. Even assuming that reionization-epoch galaxies are substantially more efficient at producing ionizing photons than their lower-redshift counterparts ($\log_{10} \xi_{\rm ion} /\text{(erg/Hz)}^{-1} = 25.5$), and integrating the UV luminosity function down to extremely faint magnitudes ($M_{\rm UV}\leq11$), we conservatively require an average $f_{\rm esc}>20\%$ from $z>6$ galaxies, more than three times larger than measurements in $z=3$ galaxies ($f_{\rm esc}\sim6\%$; \citealt{Pahl21}). A more conventional production rate of $\log_{10} \xi_{\rm ion} /\text{(erg/Hz)}^{-1} = 25.2$ requires an even more extreme $f_{\rm esc}>30\%$. Qualitatively, a ``photon budget crisis'' occurs because the \citetalias{Becker21} mean free path measurement at $z\sim6$ implies that ionizing photon sinks are far more numerous than previously assumed. 

Our models are nevertheless consistent with available reionization history constraints, as shown in Figure~\ref{fig:reion}, and with measurements of the UV LF and stellar mass density (see \S~\ref{sec:gal}). Astrophysical constraints on the ionized fraction at $z \sim 7$ (Fig.~\ref{fig:reion}, right) may point towards somewhat more rapid reionization, but our models have more gradual histories because IGM absorption becomes increasingly important later in the process.

Future observations will help resolve, or further exacerbate, this ionizing photon budget deficit. Tighter constraints on the mean free path at $z\gtrsim5.5$, with improved quasar bias modeling, would most directly improve the constraints. Improved upper limits on the dark pixel fraction at $z\sim6$ would provide a more robust limit on the IGM neutral fraction. Deeper observations of high-redshift galaxies with \emph{JWST} will deliver precise UV LFs to fainter magnitudes, potentially locating a ``turnover'' corresponding to the onset of star formation in low mass halos. The depth and wavelength coverage of \emph{JWST} spectroscopy will enable direct measurements of $\xi_{\rm ion}$ (and possibly $f_{\rm esc}$) from nebular lines, and test for consistency with our models.

\acknowledgments

We thank Joseph Hennawi and the ENIGMA group at UCSB/Leiden for access to the computing resources used in this work.

SEIB acknowledges funding from the European Research Council (ERC) under the European Union's Horizon $2020$ research and innovation programme (grant agreement No.~$740246$ ``Cosmic Gas''). 

This work was supported by the National Science Foundation through award AST-1812458. In addition, this work was directly supported by the NASA Solar System Exploration Research Virtual Institute cooperative agreement number 80ARC017M0006. We also acknowledge a NASA contract supporting the ``WFIRST Extragalactic Potential Observations (EXPO) Science Investigation Team" (15-WFIRST15-0004), administered by GSFC. This material is based upon work supported by the National Science Foundation under Grant Nos.~1636646 and 1836019, the Gordon and Betty Moore Foundation, and institutional support from the HERA collaboration partners.

\bibliographystyle{aasjournal}
 \newcommand{\noop}[1]{}

\end{document}